# A Domain Translation Framework with an Adversarial Denoising Diffusion Model to Generate Synthetic Datasets of Echocardiography Images

Technical Report

**Cristiana Tiago**
GE Vingmed Ultrasound - GE Healthcare
Horten, Norway
cristiana.tiago@gehealthcare.com

**Sten Roar Snare**
GE Vingmed Ultrasound - GE Healthcare
Horten, Norway

**Jurica Sprem**
GE Vingmed Ultrasound - GE Healthcare
Horten, Norway

**Kristin McLeod**
GE Vingmed Ultrasound - GE Healthcare
Horten, Norway

## Abstract

Currently, medical image domain translation operations show a high demand from researchers and clinicians. Amongst other capabilities, this task allows the generation of new medical images with sufficiently high image quality, making them clinically relevant. Deep Learning (DL) architectures, most specifically deep generative models, are widely used to generate and translate images from one domain to another. The proposed framework relies on an adversarial Denoising Diffusion Model (DDM) to synthesize echocardiography images and perform domain translation. Contrary to Generative Adversarial Networks (GANs), DDMs are able to generate high quality image samples with a large diversity. If a DDM is combined with a GAN, this ability to generate new data is completed at an even faster sampling time. In this work we trained an adversarial DDM combined with a GAN to learn the reverse denoising process, relying on a guide image, making sure relevant anatomical structures of each echocardiography image were kept and represented on the generated image samples. For several domain translation operations, the results verified that such generative model was able to synthesize high quality image samples: MSE: 11.50 ± 3.69, PSNR (dB): 30.48 ± 0.09, SSIM: 0.47 ± 0.03. The proposed method showed high generalization ability, introducing a framework to create echocardiography images suitable to be used for clinical research purposes.

## 1 Introduction

Echocardiography is the application of ultrasound imaging to the heart. This imaging modality is the most frequently used to image this organ, because it carries several advantages: there's a relative low cost and the equipment is portable, in comparison with Computed Tomography (CT) and Magnetic Resonance (MR). Ultrasound imaging also has the benefit of not using any ionizing radiation, this way not being harmful to the patient.

One other big advantage of echocardiography is its temporal resolution. When investigating cardiac motion, this modality still holds an advantage over others. Its wide usage in clinical practice and workflows make echocardiography a first port of call to detect pathological cases and assess the anatomy and function of the heart.

To optimize treatment pathways and spare clinicians' time to go over more severe cases, DL in healthcare has been proving its utility during the past years [Gandhi and Gandhi, 2022]. Besides these mentioned advantages, DL helps clinicians to reach the final diagnosis quicker without compromising the confidence level of it [Aljuaid and Anwar, 2022], reaching human-level performance [Scheetz et al., 2021].

In fact, DL has many and varied applications in the medical imaging domain. Image classification, anatomical structures segmentation and even detection of regions of interest are some of the most common usages of these mathematical methods. However, more recently other applications have been gaining terrain such as image generation [DuMont Schütte et al., 2021] and image domain translation/adaptation [Wang et al., 2019], which help extend the usability in this domain, where there are increasing challenges in collecting sufficient and variable datasets.

DL algorithms learn functions and patterns from data, either from time series or images. Even though with echocardiography being such a widely used cardiac imaging modality, the access to medical image data became more complicated due to all the current anonymization and privacy regulations. Consequently, there is a current need for medical data specially to train DL algorithms.

Several studies, including [Thorstensen et al., 2010] and [Uzunova et al., 2020] showed that synthetically generated images have a positive influence in the research and development of DL algorithms. Adding synthetic data to datasets made of real images adds variety to these medical image datasets and presents a solution to data scarcity, a very real phenomenon existent in the medical DL field.

Generating synthetic data using deep generative models [Ruthotto and Haber, 2021] provides a solution for this issue. Deep generative models are a subset of DL architectures trained to synthesize data. Within this group of neural networks, current methods include Variational Autoencoders and GANs. More recently, DDMs are also found under this category.

A GAN is a generative model based on a generator and a discriminator, where the former attempts to deceive the latter by minimizing the difference between the synthetically generated images and the real ones.

Diffusion models, similarly to all deep generative models, attempt to learn, by approximation, the probability distribution function representative of some training dataset. Particularly for these models, making them distinct from the rest, the generative procedure is based on the destruction of the input image by adding Gaussian noise to it, during a large enough number of steps, and consequently learn how to reverse these steps [Ho et al., 2020]. This way, it is possible to generate a synthetic image simply by denoising an initial randomly noisy input image. These models provide high fidelity/quality synthetic samples.

Creating a data augmentation tool to generate realistic echocardiography images is of need as it provides a solution to the scarcity of medical data.

### 1.1 State of the Art

Several image synthesis approaches are in practice today, with the choice of approach depending on the type of image being generated. When it comes to medical image synthesis, the choice of the imaging modality has a large impact on the selected models used to generate these images.

Most of the recent results and approaches adopt DL models to perform domain translation, with GANs being widely used since they can generate high quality samples, with a high level of realism [Dar et al., 2019], across several medical imaging modalities such as MRI [Li et al., 2019] - [Abbasi-Sureshjani et al., 2020], CT [Selim et al., 2020], and Ultrasound, namely echocardiography [Tiago et al., 2022] - [Gilbert et al., 2021], with a fast sampling time. In a GAN, the generator tries to synthesize a sample that matches the target domain, which has an inherent data distribution function. The discriminator compares this synthesized image with the ones from the training dataset in order to distinguish them.

Echocardiography raises more challenges, when compared with other imaging modalities, due to the physics behind the acquisition and image reconstruction processes. Particularly [Tiago et al., 2022] and [Gilbert et al., 2021] focused on generating 3D and 2D echocardiography, respectively. This type of medical image has inherent characteristics that strongly influence the final acquired image, namely the speckle pattern, the scanner functional characteristics, the patient's anatomy, and the sonographer's skills. Nevertheless, both works use GANs to synthesize the images, but the former considers a supervised GAN training and the latter an unsupervised approach.

However, GANs do not have a large diversity in the type of images they can generate, often leading the discriminator to converge too soon in training or to mode collapse [Isola et al., 2017]. This phenomenon is very common when training GANs which drives the model to generate image samples with less quality and very little or even no variability at all.

DDMs, on the other hand, are capable of generating samples with a large variability without compromising its high quality [Dhariwal and Nichol, 2021]. These models were initially introduced by [Sohl-Dickstein et al.] in order to save time when sampling data from a training dataset, without having to learn a great number of training steps and parameters. These models destroy the input data distribution during a sufficiently large number of time steps and then use a neural network to learn how to reverse this process, restructuring the data.

A Domain Translation Framework with an Adversarial Denoising Diffusion Model to Generate Synthetic Datasets of
Echocardiography Images                                                           TECHNICAL REPORTIn recent years, [Ho et al., 2020] and [Song et al., 2021a] attempted to show an equivalence relationship between DDMs and score based generative models, which attribute a score to probability distributions based on the likelihood of data [Song and Ermon, 2020]. The work on training DDMs, based on original statistical physics theory, showed good results both in terms of synthetic image variability [Song et al., 2021b] and also of sample quality. [Dhariwal and Nichol, 2021] demonstrated that DDMs are capable of outperforming GANs in terms of generated image quality. Furthermore, [Nichol and Dhariwal, 2021] also showed that DDMs generate images with high likelihood values when such models are trained on datasets with a wide variety of images, what brings more complexity to the training dataset probability distribution.

To tackle the longer sampling time inherent to DDMs, both [Nichol and Dhariwal, 2021] and [Song et al., 2020] presented contributions in terms of accelerating the forward diffusion process and adding noise to the input image over less steps. This way reducing the complexity of learning the reverse diffusion process and allowing to denoise image samples in a faster way, without compromising the image quality.

DDMs' application to medical image generation is yet not fully explored mainly due to its larger sampling time [Ho et al., 2020]. More recently, [Xiao et al., 2021] proposed to merge DDMs with GANs, in an attempt to make use of both generative models' strengths and tackle their individual weaknesses. This group proposed a denoising diffusion GAN, using a conditional GAN to model larger denoising steps during the reverse diffusion process.

Following the learning of conditional diffusion processes, [Özbey et al., 2023] proposed an adversarial diffusion model, SynDiff, where images from a source domain are used during training to guide the denoising diffusion process. This group applied such a model to perform medical image translation between brain MRI T1 and T2 weighted images. They were able to generate images of each domain, having an image from the other domain as a guide during the reverse diffusion process.

Taking the application of DDMs to extra dimensions, [Kim and Ye, 2022] added a deformation module to the diffusion one and attempted to generate temporal volume images (3D + time) cardiac MRI images.

In this proposed work, we applied such deep generative models to generate echocardiography images. To keep a wide variety in the generated samples and decreasing the sampling time, without compromising the image quality, we propose a data augmentation tool based on a DDM and a GAN. The proposed adversarial diffusion model generates synthetic echocardiography images and uses a GAN to learn the denoising process, whose performance is conditioned by anatomical masks of the heart. This way, these gray level masks guide the reverse diffusion process in order to maintain the anatomical information on the synthetic image. To the best of our knowledge, no previous work has presented reproducible results when it comes to generate such images using DDMs.

### 1.2 Summary of Contributions

We propose a data augmentation method to synthetically generate echocardiography images, using anatomical masks of the heart to guide the model during the image synthesis. These images are possible to use for research purposes in the medical image domain, such as the development of DL analysis tasks.

In summary and beyond the current state of the art, the main contributions of the proposed approach are:

1. The training of an adversarial diffusion model based on a DDM and a GAN, to generate synthetic echocardiography images.
2. The association of anatomical masks of the heart to the synthetically generated echocardiography image samples. This way, we tackle the lack of publicly available datasets, with labels, of echocardiography images.
3. The generation of echocardiography datasets belonging to different domains, such as from different scanners, using the proposed method to perform image domain translation.

## 2 Methodology

Fig. 1 illustrates the proposed approach. It is described in further detail in the following sections. Section II-A covers all the data collection and pre-processing steps, and Section II-B describes in further detail the working principle of the DDM and GAN behind the proposed adversarial diffusion model. Section II-C focuses on the creation of different image datasets and in Section II-D the image quality comparison metrics considered in this study are described.

### 2.1 Data Collection

The proposed adversarial diffusion model was trained on an already existing dataset of echocardiography images.

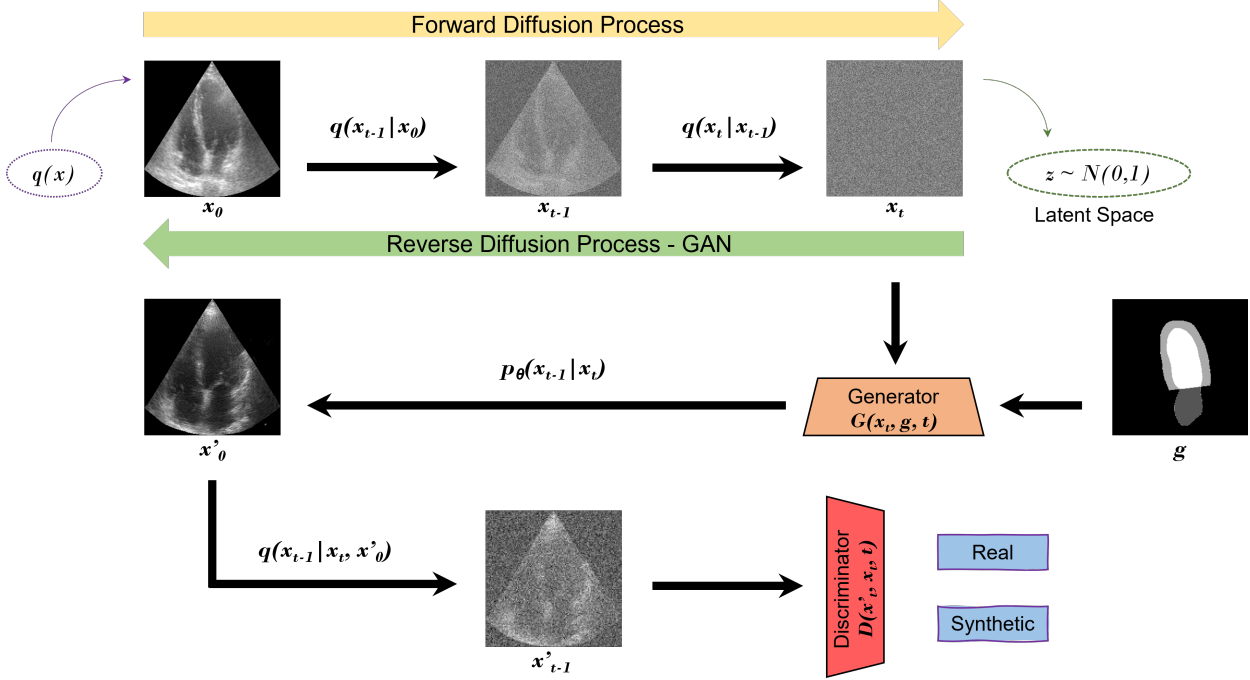

Figure 1: Proposed pipeline to generate synthetic echocardiography images from a DDM and a GAN. Forward diffusion process: during this stage, the DDM module progressively adds Gaussian noise to the training image, $x_0$, belonging to the training dataset with a distribution q(x), until a noisy image, $x_t$, is obtained after t time steps. This process creates a latent space, z, with a Gaussian distribution. The reverse diffusion process relies on a GAN to learn the reverse distribution, $p_\theta(x_t)$, and generate synthetic images, $x'_0$, in a conditional fashion.

The CAMUS dataset, proposed by [Leclerc et al., 2019], includes 2D apical two and four chamber images, acquired at end-diastole (ED) and end-systole (ES) time instances of the cardiac cycle, with poor, good, and medium image quality levels. All images were acquired with a GE Vivid E95 ultrasound scanner. For our work, we selected only ED apical four chamber (4 CH) images with all levels of image quality, resulting in 450 images, resized to 256 x 256 pixels. All the images have associated anatomical masks for the Left Ventricle (LV), Myocardium (MYO), and Left Atrium (LA). The dataset was split to train and validation sets by 90% and 10%, respectively.

Five other datasets were used for inference, to perform domain translation. All these were made of 256 x 256 apical 4CH images and included anatomical masks with the same structures considered in the CAMUS dataset. Table 1 summarizes all the considered datasets in this work.

Table 1: Summary of all used datasets in this work.

| Dataset | Origin | Acquisition scanner | Labels | Original image size | Final image size | Stage |
|---|---|---|---|---|---|---|
| CAMUS | Publicly available | GE Vivid E95 | LA, LV, and MYO | Variable | 6*256 x 256 | DDM Training |
| Vscan | GE Healthcare | GE Vscan Extend | * | 416 x 240 | | Inference (domain translation) |
| Vscan Air | GE Healthcare | GE Vscan Air | * | 2040 x 1024 | | Inference (domain translation) |
| EchoNet | Publicly available | Multiple Philips and Siemens scanners | LV | 112 x 112 | | Inference (domain translation) |
| 2D + t | GE Healthcare | Multiple GE Vivid models (except E95) | * | 1016 x 708 | | Inference (domain translation) |
| 3D (spatial) | GE Healthcare | Multiple GE Vivid models (except E95) | * | Variable | | Inference (domain translation) |

* Labels for the LA, LV, and MYO were created for these datasets.

First, the EchoNet-Dynamic dataset presented by [Ouyang et al., 2020], was used. This dataset contains more than ten thousand labeled echocardiogram videos. For the task of generating synthetic echocardiography images, only the ED frames of the echocardiographic videos were used. The anatomical masks associated with this dataset only showed the LV area. We then added the MYO and LA areas to them.

A second dataset was also made up of ED frames extracted from 3D (3 spatial dimensions) echocardiography images, acquired with different GE Vivid ultrasound scanners.

Two other datasets were also created using another two handheld GE ultrasound scanners: the Vscan Extend and the Vscan Air. The former is a pocket-sized scanner, and the latter is used to image the heart using a wireless probe and displaying the image on a smartphone. We created the anatomical labels for the second, third and fourth datasets.

A fifth dataset included ED frames extracted from 2D + time (2D + t) images, all of them acquired with GE Vivid ultrasound scanners, different from the GE Vivid E95. This one was previously labeled by a cardiologist.

## 2.2 Adversarial Diffusion Model Training

The mathematical reasoning behind diffusion models was initially proposed by [Sohl-Dickstein et al.]. This group showed that it is possible to reconstruct a noisy image in order to generate a sample belonging to a certain dataset with a defined probability distribution function.

The denoising principle behind shows that these generative models offer higher quality and more variate image samples than others.

By themselves, DDMs are known to be based on unconditional diffusion processes applied during a large number of steps. However, the proposed adversarial diffusion model performs the reverse diffusion process in a conditional fashion. In order to synthesize image samples with similar statistical properties as the training dataset, the adversarial model uses images from a second domain to guide, i.e. condition, the reverse denoising algorithm. Furthermore, our adversarial DDM learns a faster reverse diffusion process which has a large step size instead of several small denoising instants.

As represented on Fig. 1, the CAMUS dataset described on the previous section was used to train the proposed adversarial diffusion model. During the forward process of the training, an image $x_0$ is sampled from the training dataset with a probability distribution $q(x)$ and Gaussian noise is added to the image sample over $T$ time steps. This process creates a Markov chain with a pre-defined variance $\beta_t$, defining the forward data distribution as:

$$q(x_t|x_{t-1}) = \mathcal{N}(x_t; \sqrt{1-\beta_t}x_{t-1}, \beta_t \mathbf{I}) \tag{1}$$

However, since the adversarial scenario allows the definition of a large step size, reducing the total number of denoising steps to be learned, the forward process can be re-written as:

$$q(x_t|x_{t-k}) = \mathcal{N}(x_t; \sqrt{1-\beta_t}x_{t-k}, \beta_t \mathbf{I}) \tag{2}$$

where $k$ is the step size and $k >> 1$, as defined in [Özbey et al., 2023].

On the other hand, the reverse denoising process is also a Markov chain approximated by a Gaussian distribution $p_\theta(x_{0:T})$, where $\theta$ are the predicted parameters of the reverse diffusion probability distribution, estimated by the GAN:

$$p_\theta(x_{t-k}|x_t) = \mathcal{N}(x_{t-k}; \mu_\theta(x_t, t), \sigma_t^2 \mathbf{I}) \tag{3}$$

The training process of our adversarial DDM aims to minimize the difference between the conditional GAN predicted probability distribution $p_\theta$, and the original training distribution $q(x)$:

$$\min_\theta \mathcal{L} = \min_\theta \sum_{t>=1} \mathbb{E}_{q(x_t)}\left[D(q(x_{t-k}|x_t)||p_\theta(x_{t-k}|x_t))\right] \tag{4}$$

where $D$ represents the Kullback-Leibler divergence used in this loss function [Ho et al., 2020].

In the proposed architecture $x'_0$ is reconstructed by the generator of the GAN from the latent space $z$, where the feature information about the training data is encoded, and which follows a normal distribution.

Associated with the echocardiography images from the CAMUS dataset, there are anatomical masks which were used to guide the denoising process. This way, the GAN performance is conditioned when estimating the denoising distribution $p_\theta$.

Given a source image $y$ to guide the reverse diffusion process, the generator $G$ attempts to estimate $p_\theta(x_{t-k}|x_t, y)$ by synthesizing $x'_{t-k}$ such that $x'_{t-k} \sim p_\theta(x_{t-k}|x_t, y)$. The discriminator $D(x'_{t-k}, x_t, t)$ distinguishes between samples from either the real probability distribution, $q(x)$, or the predicted $p_\theta(x)$.

### 2.3 Domain Translation - Inference

Domain translation allows to transform images from a domain A to a domain B, so that the generated, i.e. domain-translated, images have similar characteristics to the ones belonging to the initial domain [Murez et al., 2018]. This operation learns how to do such translation by analyzing the probability distribution of the initial dataset and iteratively compare it with the statistical distribution of the target domain [Zhu et al., 2020].

After training the adversarial diffusion model using the CAMUS dataset, at inference time, the datasets described on Section II-A were considered as input to the trained model. These inference steps allowed to perform domain translation and create synthetic datasets with characteristics similar to CAMUS.

### 2.4 Image Quality Comparison Metrics

To evaluate the quality of the generated image samples from all different synthesized datasets described before, several image quality metrics were calculated and compared.

The most commonly used image quality estimator is the Mean Squared Error (MSE), which quantifies the difference between two different images, measuring the differences pixel by pixel. If the synthetic image is similar to the ground truth one, then this error will be low.

The Peak Signal-to-Noise (PSNR) ratio takes into account the signal from the original image and the noise, i.e. error, of the generated sample. This metric is presented in dB and [S. Faragallah et al., 2020] considers values around and above 30 dB as representing good quality synthetic image samples.

Both these metrics are pixel based. To evaluate the quality of generated images using a method more similar to the human visual system, the Structural Similarity Method (SSIM) [Renieblas et al., 2017] was considered. SSIM takes into account the preserved and changed edges information between the original image and the generated one, and also the texture differences. This index takes values between 0 and 1, with higher values reflecting a larger image similarity.

Specifically created to measure the performance of GANs, the Fréchet Inception Distance (FID) was defined by [Heusel et al., 2017] to evaluate the quality of the generated samples from different datasets. Contrary to the already described metrics, the FID score does not directly compare generated and real images, but it measures the distance between the statistical distribution of synthetic and real datasets [Skandarani et al., 2023]. The lower this score is, the smaller the difference between the datasets.

## 3 Results

The training parameters and training time of the proposed adversarial diffusion models are described in Section III-A, and Section III-B details the results of the domain translation operation, together with the image quality comparison metrics obtained.

### 3.1 Adversarial Diffusion Model Training

The proposed adversarial DDM was trained during 500 epochs and for a total of four diffusion steps. The upper and lower bounds for the variance of the predicted distribution were kept the same as in [Özbey et al., 2023]. The model was built using PyTorch [Paszke et al.] and it was trained on a computer equipped with four NVIDIA GeForce RTX 2080 GPUs (multiple GPU training). Training took approximately forty hours.

Fig. 2 gives an overview of the training results during the validation steps. It shows a generated sample with similar characteristics to the images in the training dataset, and keeping the anatomical information present in the guide image.

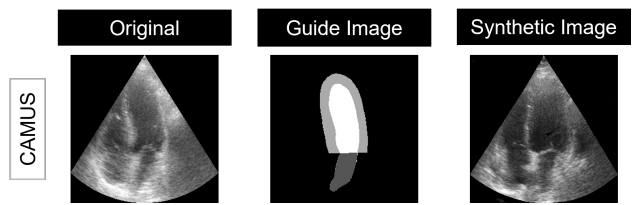

Figure 2: Adversarial diffusion model training results. For the validation image shown on the left, the image on the right is the generated sample, outputted by the trained model.

### 3.2 Domain Translation - Comparison Metrics

After training the adversarial diffusion model, it was used to perform domain translation operations. For each of the five previously created datasets with different image characteristics, a synthetic dataset with properties similar to CAMUS was generated.

Fig. 3 shows the best generated image sample from each of the domain translation operations performed.

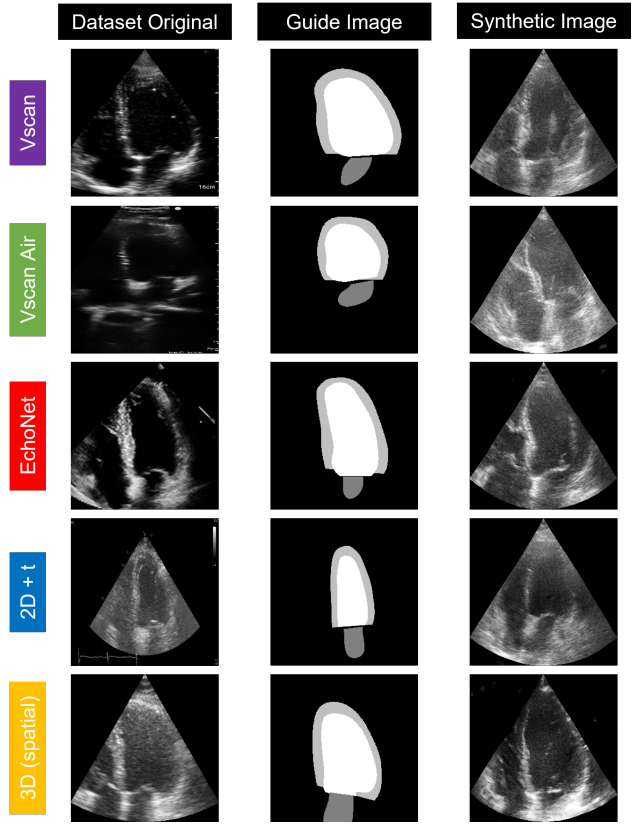

Figure 3: Domain translation results. Best generated image from each of the inference datasets. All the synthetic images show characteristics of the CAMUS dataset and keep the anatomical information present in the guide image (white area – LV, dark gray area – LA, light gray area – MYO).

The FID score, in Table 2, gives the overview of the complete dataset quality, instead of comparing individual image samples. Fig. 4 shows examples of the worst, median, and best generated images, in terms of the PSNR value, after the domain translation operation.

Table 3 lists the image comparison metrics calculated between the generated sample and the ground truth image, for each test image belonging to the inference datasets.

Table 2: FID scores for each original inference dataset (before domain translation) and each synthetic dataset (after domain translation), compared with the training CAMUS dataset. The best scores are highlighted.

|  | Inference Datasets (before domain translation) | | | | | Synthetic Datasets (after domain translation) | | | | |
| --- | --- | --- | --- | --- | --- | --- | --- | --- | --- | --- |
|  | Vscan | Vscan Air | EchoNet | 2D + t | 3D (spatial) | Vscan | Vscan Air | EchoNet | 2D + t | 3D (spatial) |
| FID | 279.53 | 332.73 | 260.42 | 189.55 | **61.08** | 70.18 | 81.22 | 60.17 | 79.28 | **50.87** |

## 4 Discussion

The proposed adversarial diffusion model architecture, based on a DDM and a GAN, proved to be able to produce a wide variety of generated image samples with a fast sampling time. In fact, training such a complex model took less than two days. This result was expected since diffusion models were conceived to learn less training parameters, in comparison to other deep generative models such as GANs, making the training lighter and faster without compromising the final output quality.

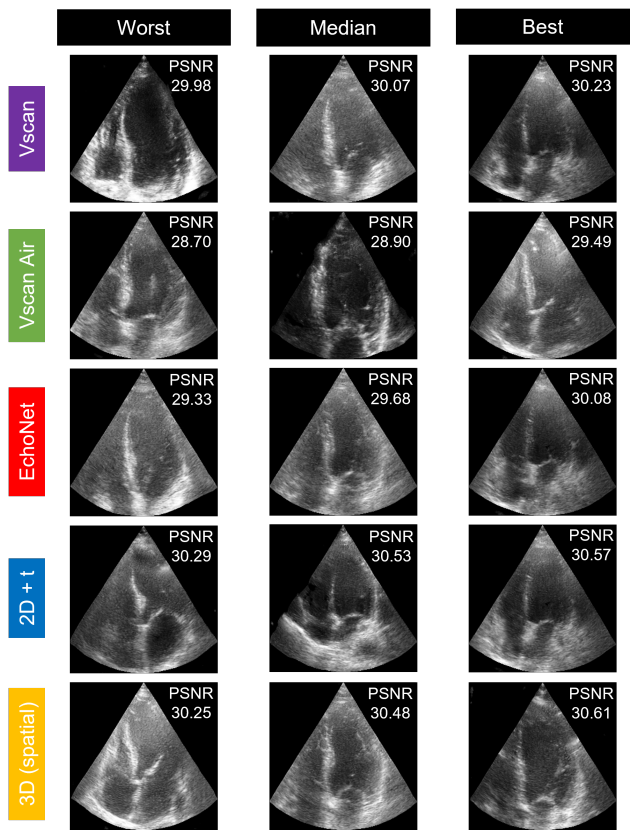

Figure 4: Worst, median, and best synthesized image from each of the inference datasets, in terms of PSNR. The worst images are not totally discrepant from the best ones.

Table 3: Comparison metrics (average ± standard deviation) for the domain translation operations. MSE, PSNR (dB), and SSIM were calculated for all the images in the 5 inference datasets. The best scores are highlighted.

|  | Metrics | | |
|---|---|---|---|
|  | MSE | PSNR (dB) | SSIM |
| Vscan | $18.27 \pm 9.26$ | $30.09 \pm 0.12$ | $0.37 \pm 0.01$ |
| Vscan Air | $30.60 \pm 7.79$ | $28.94 \pm 0.30$ | $0.18 \pm 0.03$ |
| EchoNet | $22.39 \pm 5.41$ | $29.65 \pm 0.20$ | $0.31 \pm 0.02$ |
| 2D + t | $11.95 \pm 7.21$ | $30.48 \pm 0.13$ | $0.40 \pm 0.01$ |
| 3D (spatial) | $\mathbf{11.50 \pm 3.69}$ | $\mathbf{30.48 \pm 0.09}$ | $\mathbf{0.47 \pm 0.03}$ |

None of the generated image samples required post-processing operations, for example to fix the cone shape [Tiago et al., 2022] or remove unwanted noise, in opposition to what has been reported when generating images with other deep generative models, where these operations are often required. DDMs hold this advantage of generating more visually accurate image samples without requiring additional post-processing steps [Dhariwal and Nichol, 2021]. On the contrary to what happens with GANs, the image samples generated via the adversarial DDM show no artifacts.

After collecting data from five different echocardiography datasets with different image characteristics amongst them, the trained model was then used to perform different tasks of image domain translation.

In terms of image acquisition, the echocardiography scans acquired with the Vscan Air (Fig. 3) are substantially different from the images one would get if the GE Vivid E95 would be used, due to the nature of the ultrasound probe used by the former scanner. This dataset characteristic is supported by the results on Table 2, where the FID score, for the Vscan Air dataset is the highest amongst all the inference datasets, when compared to CAMUS, reflecting this difference.

From Table 2 it is also visible that the inference dataset containing 2D apical 4CH echocardiography images extracted from 3D scans (where the 3 spatial dimensions were considered), 3D (spatial), is the most similar dataset to the CAMUS dataset, amongst all the five inference datasets, as it holds the lowest FID score. Consequent manual inspection confirmed that these datasets are visually the most similar.

Five domain translation operations were performed and shown in this work. During each of these, the trained adversarial diffusion model generated an image sample corresponding to each image in the considered dataset. The generated images were then compared to the ground truth and the MSE, PSNR, and SSIM were calculated (Table 3).

The 3D (spatial) dataset showed the best results for all these three metrics. The high value of the PSNR indicates that the information present in the original inference images is preserved and visible on the synthetic images generated with the adversarial diffusion model.

The SSIM value for the Vscan Air dataset holds the lowest value reinforcing the conclusion described earlier, stating that this dataset images belong to a domain which is the most different from the CAMUS images domain. On the other hand, a PSNR close to 30 dB reflects that the domain translation operation was still able to synthesize images with meaningful information encoded on them.

After the domain translation operations, the FID score was calculated for each of the synthetic datasets (Table 2). The 3D (spatial) synthetic dataset is still the one registering the lowest FID value amongst all the synthetic datasets. The difference between the FID scores obtained before and after the domain translation operations are indicative of the generalization ability of the proposed adversarial diffusion model. Table 2 shows a significant decrease in all datasets FID scores, after domain translation. The scores represent a smaller difference between the probability distribution of each synthetic dataset and the CAMUS. The EchoNet synthetic dataset has a smaller FID score than the 2D + t, even though, before domain translation, the opposite scenario, i.e. smaller FID for the 2D + t dataset, was verified.

The adversarial DDM trained was able to generate variate samples, closely depicting the LA, LV, and MYO, present on apical 4CH echocardiography images (Fig. 4). The images considered as worst, in terms of PSNR, still illustrate these structures and are not completely divergent from the best ones.

Presented results described and discussed in this section support the initial premises; namely that diffusion models are lighter and quicker to train and are able to generate high quality image samples. Creating an adversarial diffusion model, by using a GAN to learn the reverse diffusion process, brings the advantage of generating images with a small sampling time. The developed approach can be used to generate synthetic datasets of echocardiography image samples and also improve the quality of lower-resolution ones. This way, the adversarial DDM is a resource to generate images belonging to different image domains, helping in the development of DL models that perform equally well irrespective of the imaging scanner/vendor.

To the best of our knowledge, diffusion models were not yet used to generate clinically relevant echocardiography images, nor used to perform domain translation operations between substantially different medical image datasets. Our work demonstrated that such tasks are possible and the generated echocardiography images have high quality and include meaningful anatomical information, since anatomical masks were used to guide the reverse diffusion process. In the future, the influence of the type of guide image used during the adversarial learning process will be further explored. Also, the analysis of the synthetic images in the clinical scenario will be assessed, by working closely with clinical end-users.

## 5 Conclusion

A domain translation framework based on an adversarial diffusion model was proposed, in order to generate synthetic datasets of echocardiography images. In the medical scenario, DL approaches outperform other methods for some tasks, including medical image generation and domain translation operations. These DL methods, however, require a large amount of data during their training and development.

The proposed framework relies on the usage of state of the art models and methods to both generate echocardiography images and also perform domain translation. These tasks allow to create a large amount of variate medical image data with clinical relevance which can be used for research and learning methodologies.

Furthermore, the proposed model showed a great generalization capacity, being able to synthesize echocardiography images with a large variability.

## Acknowledgements

This work was supported by the European Union's Horizon 2020 research and innovation programme under the Marie – Skłodowska – Curie grant agreement No 860745.